\documentclass[runningheads, eepic, 10pt]{llncs}
\usepackage{amssymb, amsmath}

\begin{document}

\pagestyle{headings}

\mainmatter

\title{Stream cipher based on quasigroup string transformations in $\bbbz_p^*$}

\titlerunning{A quasigroup stream cipher in $\bbbz_p^*$}

\author{D. Gligoroski \inst{1}\\ First version: 9 December
2003\\
First revision: 30 March 2004\\
Second revision: 22 April 2004}

\authorrunning{Gligoroski}

\institute{$^1$University ``St. Cyril and Methodious'',
 \\Faculty of
 Natural Sciences, \\ Institute of Informatics, P. O. Box 162, \\
 Skopje, Republic
 of Macedonia\\
 \email{gligoroski@yahoo.com}
 }
\maketitle

\begin{abstract}
In this paper we design a stream cipher that uses the algebraic
structure of the multiplicative group $\bbbz_p^*$ (where p is a
big prime number used in ElGamal algorithm), by defining a
quasigroup of order $p-1$ and by doing quasigroup string
transformations. The cryptographical strength of the proposed
stream cipher is based on the fact that breaking it would be at
least as hard as solving systems of multivariate polynomial
equations modulo big prime number $p$ which is NP-hard problem and
there are no known fast randomized or deterministic algorithms for
solving it. Unlikely the speed of known ciphers that work in
$\bbbz_p^*$ for big prime numbers $p$, the speed of this stream
cipher both in encryption and decryption phase is comparable with
the fastest symmetric-key stream ciphers.

\vspace{0.5cm} {\it Key words: quasigroups, quasigroup string
transformations, stream cipher, public-key, ElGamal}

\vspace{0.5cm} {\it AMS Mathematics Subject Classification
(2000)}: 94A60, 94A62, 68P30

\end{abstract}

\section{Introduction}%

From the point of view how encryption algorithms encrypt
information that is repeated several times during the phase of
communication, they are divided on stream ciphers and block
ciphers. While block ciphers always give the same output of cipher
texts for the same input blocks of plain text, the stream ciphers
give different outputs for the same sequences of plain text. On
the other side, depending on the type of the keys used in
cryptographic algorithm, and the way the keys are used, there is
another classification of encryption algorithms: symmetric-key and
public-key algorithms. Symmetric-key algorithms need the
correspondents in the communication to share a same key that is
previously exchanged through some secure channel that is out of
the scope of the definition of the algorithm, while in public-key
algorithms the problem of exchanging the communication key is a
part of the algorithm and no secure channel is necessary for that
purpose.

Stream cipher algorithms can be either symmetric-key or
public-key. Regarding the speed of encryption and decryption
procedures, symmetric-key stream ciphers are much more faster then
public-key ones. That is because the symmetric-key stream ciphers
usually use fast register operations such as shifting, rotation,
and bit by bit logical operations, while the most popular and
known public-key algorithms usually use modular exponentiation.
Thus, the public-key algorithms are around 1000 times slower then
symmetric-key algorithms.

A well known public-key stream cipher is Blum-Goldwasser
probabilistic public-key encryption scheme \cite{BlumGoldwaser}.
Even though the speed of that algorithm in encryption phase is
much faster then RSA encryption, the speed of that algorithm in
decryption phase is similar or in some cases even slower then the
speed of RSA algorithm, (\cite{HAC} p. 310-311). In fact the lack
of the speed of public-key stream ciphers is one of the main
reasons why they are not widely used in stream communication.

Diffie-Hellman algorithm was proposed in 1976
\cite{DiffieHellman} and introduced the concept of public-key
cryptography. That algorithm usually is used for establishing a
key exchange between two correspondents, and then, the
communication is usually continued by some symmetric fast
algorithm (either block or stream cipher). In 1985 ElGamal
proposed a public-key cryptosystem based on Diffie-Hellman
algorithm \cite{ElGamal}. One of the disadvantages of ElGamal
algorithm is that cipher text is two times longer then
corresponding plain text, which makes it unsuitable for using
it as a stream cipher.

In this paper beside the theory of finite fields we use also the
theory of quasigroups and Latin Squares. Although quasigroups (or
Latin squares) are used in design of many modern symmetric
cryptographic algorithms \cite{vaudenay94}, \cite{schnorr95} they
are not in the main stream of cryptographic paradigms. During the
last 10 years several cryptographic algorithms were developed
based on quasigroups \cite{Bakhtiari97}, \cite{DenesKeedwel92},
\cite{Czeslaw}. Those algorithms base their security on
assumptions that other problems such as factoring of natural
numbers or discrete logarithm problems can not be solved in
polynomial time - and thus have solid theoretical ground for their
security. However, for all of those algorithms, because they
usually use sets of Latin squares (quasigroups), their
implementation is several orders of magnitude slower than other
cryptographic algorithms in their category, based usually on bit
manipulation and shifting registers.

Excellent introductory materials about
theory of quasigroups the reader can find in \cite{Belousov} and \cite{DenesKeedwel81}
and some applications of quasigroups and Latin squares in
\cite{Hall}, \cite{AndreFritz}, \cite{McKay}, \cite{stinson99}.

In cryptographic algorithms introduced in \cite{LIRA} and
\cite{Quasi1}, and the following papers \cite{Quasi2},
\cite{YTalk}, \cite{Quasi12}, \cite{EdonF} and \cite{SmileQ} the
authors use quasigroups to define so-called ``quasigroup string
transformations''. By those algorithms they define a stream cipher
whose principles are used in this paper. For effective encryption
and decryption, the quasigroup stream cipher uses a set of leaders
that are in fact the secret and symmetric key. However,
quasigroups used in those algorithms are of the order from 16 to
256, and complete multiplicative table have to be known, before
encryption/decryption starts.

In the paper \cite{Marnas} in order to solve the problem of fast
generation of a quasigroups of order $p-1$ where $p$ is a prime
number, authors propose a fast way for generating a quasigroups by
knowing only the first row in the multiplicative table of the
quasigroup. That first row is in fact a permutation of the
elements $\bbbz_p^*=\bbbz_p \setminus \{0\}=\{1,2,...,p-1\}$ and
by knowing only that permutation it is possible to define the
product of any two elements such that a quasigroup will be formed.

In this paper we will define a stream cipher that in its
initialization phase uses ElGamal algorithm, then the encryption
is made by using quasigroup string transformations and the
definition of a quasigroup is based by knowing only one
permutation in the set of $\bbbz_p^*$.

The organization of the paper is following: In Section 2 we will
give basic definitions of the ElGamal algorithm, quasigroup stream
cipher and fast quasigroup definition from a known permutation. In
Section 3 we will define the new stream cipher and we will give an
example with a small value of $p$, in Section 4 we will examine
the cryptographical strength of the proposed stream cipher, and in
Sections 5 we will give the conclusions.

\section{Basic definitions}

In our description of cryptographic algorithms we will use the
usual notification that the correspondents in the communication
are Alice and Bob.

\subsection{Basic ElGamal encryption algorithm}

The ElGamal encryption algorithm uses a big prime number $p$, and
uses the operations of modular exponentiation and modular
multiplication. There are three phases of the algorithm: Key
generation, Encryption and Decryption. The algorithm is the
following:

\begin{description}
  \item[Key generation] Alice generates her public and private
  keys as follows:
\begin{itemize}
  \item [1.] Generate a large random prime number $p$ and a generator
  $\alpha$ of the multiplicative group $\bbbz_p^*$ of the integers
  $\{1,2,\ldots,p-1\}$.
  \item [2.] Select a random integer $a$, $1\le a \le p-2$ and compute
  $\alpha^a {\tt mod}\, p$.
  \item [3.] Alice's public key is the triplet $(p,\alpha,\alpha^a)$;
  Alice's private key is $a$.
\end{itemize}

  \item[Encryption] Bob encrypts a message $m$ for Alice by doing the following:
\begin{itemize}
  \item [1.] Obtain Alice's authentic public key
  $(p,\alpha,\alpha^a)$.
  \item [2.] Represent the message as an integer $m$ in the range $\{0,1,\ldots,p - 1\}$.
  \item [3.] Select a random integer $e$, $1 \le e \le p - 2$.
  \item [4.] Compute $\gamma = \alpha^e {\tt mod}\, p$ and $\delta = m
  \cdot (\alpha^a)^e {\tt mod}\, p$.
  \item [5.] Send the ciphertext $c=(\gamma,\delta)$.
\end{itemize}

  \item[Decryption] To recover the message $m$ Alice should do the
  following:
\begin{itemize}
  \item [1.] Use the private key $a$ to compute $\alpha^{-a e} = \gamma^{-a}$
  \item [2.] Recover $m$ by computing $m=\delta \cdot \alpha^{-a e}
  {\tt mod}\, p$
\end{itemize}

\end{description}

It is obvious that message expansion in ElGamal algorithm is by
factor 2, because Bob sends the cipher text $c=(\gamma,\delta)$
that has twice the length of the message $m$. That fact is
considered as a serious disadvantage of the algorithm. Simple
analysis of the algorithms show that in a phase of encryption it
uses two modular exponentiations and one modular multiplication,
while in phase od Decryption it uses one modular exponentiation,
one calculation of an inverse element in multiplicative group
$\bbbz_p^*$ (calculation of the element $\gamma^{-1} {\tt mod}\,
p$) and one modular multiplication. For the security analysis, and
security issues about used prime numbers in ElGamal algorithm the
reader can see \cite{HAC}.

\subsection{Definition of basic quasigroup string transformations}

In this subsection we will give some definitions from the theory
of quasigroups and define a basic quasigroup string
transformations. We say ``basic'' string transformations, because
in \cite{Quasi1} much more complex quasigroup string
transformations are defined, but we will not use them in our
definition of the stream cipher.

\begin{definition}
Let $Q=\{a_1, a_2, \dots, a_n\}$ be a finite set of $n$ elements.
A quasigroup $(Q,*)$ is a groupoid satisfying the law
\begin{equation}(\forall u,v\in Q)(\exists!x,y\in Q) \quad u*x = v\ \& \ y*u
= v. \label{QuasigroupDef}\end{equation} \label{D1}
\end{definition}

\vspace{-0.7cm} Given a quasigroup $(Q,*)$ a new operation
$*^{-1}$ on the set $Q$ can be derived by: \begin{equation}
*^{-1}(x,y) =z \Longleftrightarrow  x*z=y \label{2} \end{equation}

It easy to prove the following
\begin{lemma}
The groupoid $(Q,*^{-1})$ is a quasigroup.
 \qed
\end{lemma}

Instead of the symbol $*^{-1}$ we will use the symbol $\setminus$
and we will say that the quasigroup $(Q,\setminus)$ is the left
parastrophe (or conjugate in some literature) adjoint to the
quasigroup $(Q,*)$.

Then from the definition of $\setminus$ it follows that
\begin{equation} x*y = z\ \Longleftrightarrow\ y=x\setminus z\
. \label{3}
\end{equation}
and
\begin{equation}
x\setminus (x*y) =y,\ x*(x\setminus y)=y. \label{4}
\end{equation}

In what follow we will give basic definitions for quasigroup
string transformations and address several theorems and properties
which are proved in \cite{Quasi1}.

Consider an alphabet (i.e. a finite set) $Q$, and denote by $Q^+$
the set of all nonempty words (i.e. finite strings) formed by the
elements of $Q$. The elements of $Q^+$ will be rather denoted by
$a_1a_2\dots a_n$ than $(a_1,a_2,\dots,a_n)$, where $a_i\in Q$.
Let $*$ be a quasigroup operation on  the set $Q$, i.e. consider a
quasigroup $(Q,*)$. For each $a\in Q$ we define two functions
$e_a, d_a:Q^+ \longrightarrow Q^+$ as follows.

Let $a_i\in Q, \ \alpha=a_1a_2\dots a_n$. Then
$$e_a(\alpha)= b_1b_2\dots b_n \Longleftrightarrow b_1=a*a_1,\
b_2=b_1*a_2,\dots,\ b_n = b_{n-1}*a_n$$ i.e. $b_{i+1}=b_i*a_{i+1}$
for each $i=0,1,\dots,n-1$, where $b_0=a,$

and
$$d_a(\alpha)= c_1c_2\dots c_n \Longleftrightarrow c_1=a*a_1,\
c_2=a_1*a_2,\dots,\ c_n = a_{n-1}*a_n$$ i.e. $c_{i+1}=a_i*a_{i+1}$
for each $i=0,1,\dots,n-1$, where $a_0=a.$

\begin{figure}[h]
\begin{center}

\unitlength=1cm
\begin{picture}(10,1.5)(-5,-0.5)
\thinlines \put(-4,0){\line(1,0){8}} \thicklines
\put(-3,-1){\line(0,1){2}} \thinlines
\multiput(-2,-1)(1,0){2}{\line(0,1){2}}
\multiput(2,-1)(1,0){3}{\line(0,1){2}}

\put(-2.5,0.5){$a_1$} \put(-1.5,0.5){$a_2$} \put(0.5,0.5){$\dots$}
\put(2.2,0.5){$a_{n-1}$} \put(3.5,0.5){$a_n$}

\put(-3.5,-0.65){$a$} \put(-2.5,-0.65){$b_1$}
\put(-1.5,-0.65){$b_2$} \put(0.5,-0.65){$\dots$}
\put(2.2,-0.65){$b_{n-1}$} \put(3.5,-0.65){$b_n$}

\multiput(-3.35,-0.5)(1,0){3}{\vector(1,1){0.85}}
\multiput(1.65,-0.5)(1,0){2}{\vector(1,1){0.85}}


\multiput(-2.33,0.35)(1,0){2}{\vector(0,-1){0.65}}
\multiput(-2.37,0.35)(1,0){2}{\vector(0,-1){0.65}}

\multiput(2.63,0.35)(1,0){2}{\vector(0,-1){0.65}}
\multiput(2.67,0.35)(1,0){2}{\vector(0,-1){0.65}}


\end{picture}
\end{center}
\caption{Graphical representation of $e_a$ function} \label{Fig1}
\end{figure}
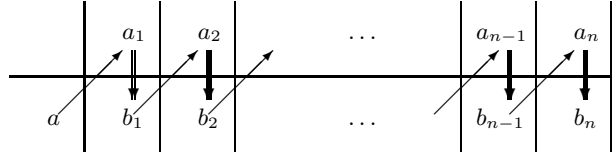

\begin{figure}[h]
\begin{center}

\unitlength=1cm
\begin{picture}(10,1.5)(-5,-0.5)
\thinlines \put(-4,0){\line(1,0){8}} \thicklines
\put(-3,-1){\line(0,1){2}} \thinlines
\multiput(-2,-1)(1,0){2}{\line(0,1){2}}
\multiput(2,-1)(1,0){3}{\line(0,1){2}}

\put(-3.5,0.5){$a$} \put(-2.5,0.5){$a_1$} \put(-1.5,0.5){$a_2$}
\put(0.5,0.5){$\dots$} \put(2.2,0.5){$a_{n-1}$}
\put(3.5,0.5){$a_n$}

\put(-2.5,-0.65){$b_1$} \put(-1.5,-0.65){$b_2$}
\put(0.5,-0.65){$\dots$} \put(2.2,-0.65){$b_{n-1}$}
\put(3.5,-0.65){$b_n$}

\put(-3.31,0.6){\vector(1,0){0.7}}
\put(-2.25,0.6){\vector(1,0){0.65}}
\put(-1.21,0.6){\vector(1,0){0.65}}
\put(1.50,0.6){\vector(1,0){0.7}}
\put(2.9,0.6){\vector(1,0){0.55}}


\multiput(-2.33,0.35)(1,0){2}{\vector(0,-1){0.65}}
\multiput(-2.37,0.35)(1,0){2}{\vector(0,-1){0.65}}

\multiput(2.63,0.35)(1,0){2}{\vector(0,-1){0.65}}
\multiput(2.67,0.35)(1,0){2}{\vector(0,-1){0.65}}


\end{picture}
\end{center}
\caption{Graphical representation of $d_a$ function} \label{Fig2}
\end{figure}
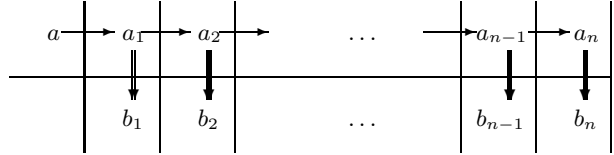

\begin{definition}
The functions $e_a,\ d_a$ are called e- and d- string
transformation of $Q^+$  based on the operation $*$  with leader
$a$.
\end{definition}

Note that $e_a \circ d_a = d_a \circ e_a = 1_a$ i.e. $e_a$ and
$d_a$ are mutually inverse string transformations. A graphical
representation of $e_a$ and $d_a$ is shown on Fig. \ref{Fig1} and
Fig. \ref{Fig2}. Next we will extend the definition of e- and d-
string transformations with the following

\begin{definition}
If we choose $k$ leaders $a_1,\ a_2,\ \dots,\ a_k\in Q$ (not
necessarily distinct), then the compositions of mappings
$$E_k = E_{a_1\dots a_k}=e_{a_1}\circ e_{a_2}\circ\dots\circ e_{a_k} $$
and
$$D_k = D_{a_1\dots a_k}=d_{a_1}\circ d_{a_2}\circ\dots\circ d_{a_k} $$
are called E- and D- quasigroup string transformations of $Q^+$
respectively.
\end{definition}

In \cite{Quasi1} the following two lemmas are proved:

\begin{lemma}
The functions $E_k$ and $D_k$ are permutations on $Q^+$. \qed
\end{lemma}

\begin{lemma}
In a quasigroup $(Q,*)$, with a given set of $k$ leaders $\{a_1,
a_2,\ldots,a_k\}$ the inverse of $E_k=E_{a_1\dots a_k} =
e_{a_1}\circ\dots\circ e_{a_k}$, is $E_k^{-1}= D_{a_k\dots\ a_1}=
d_{a_k}\circ\dots\circ\ d_{a_1}$.
\end{lemma}

Now it is clear that for any quasigroup string transformation $E$
the pair of functions $(E,E^{-1})$ can be considered as a pair of
an encryption and decryption function for the strings on an
alphabet $Q$. More formally we give the following definition of a
quasigroup stream cipher:

\begin{definition}
For a given quasigroup $(Q,*)$, and a given $k$-tuple $(a_1,
a_2,\ldots,a_k)$, of leaders $a_i \in Q$, the system $((Q,*),(a_1,
a_2,\ldots,a_k),E_{a_1\dots a_k},D_{a_k\dots a_1})$ defines a
quasigroup stream cipher on the strings in $Q^+$.
\end{definition}

\subsection{Definition of a quasigroup of big order $p-1$}

The construction of a Latin squares is discussed in
\cite{Czeslaw}, \cite{Hall}
 and \cite{McKay}. However, the construction of such
Latin squares is not suitable for our purposes in this paper,
because we want to define a quasigroup of order $p-1$ where $p$ is
big prime number with more then 1024 bits. That problem can be
solved by the approach that is described in \cite{Marnas}. Namely,
if we have a permutation $P=\left(
\begin{array}{cccccc}
1&2&\cdots&j&\cdots&p-1\\
a_{1 1}&a_{1 2}&\cdots&a_{1 j}&\cdots&a_{1 p-1}
\end{array}
\right) $, where $(a_{1 1},a_{1 2},\cdots,a_{1 j},\cdots,a_{1
p-1})$ is the first row of the quasigroup that we want to define,
then by defining $i*j=i\times a_{1\,j}\ {\tt mod}\, p$ we will
define a quasigroup $(Q,*)$ of order $p-1$.

We will define a permutation of the elements in $\bbbz_p^*$ by the
following lemma:
\begin{lemma}
For a given prime number $p$, and a given number $K, 1\le K \le
p-2$, the function $f_K(j)=\frac{1}{1+(K+j)\, {\tt mod}\, (p-1)} \
{\tt mod}\, p$ is a permutation of the elements in $\bbbz_p^*$.
\qed
\end{lemma}

Now, we can prove the following
\begin{lemma}
The multiplication operation $*$ defined in the set
$Q=\{1,2,\ldots,p-1\}$ as:
\begin{equation}
i*j=i \times f_K(j)\ {\tt mod}\, p \label{5}
\end{equation}
defines a quasigroup $(Q,*)$. \qed
\end{lemma}

From the last Lemma, we have the following
\begin{corollary}
If we define the following function \begin{equation} g(i,j,K)= ((i
\times j^{-1}\ {\tt mod}\ p) -1 -K)\ {\tt mod}\ (p-1)\label{6}
\end{equation}
that takes the arguments $i,j$ and $K$ from the set
$\{1,2,\ldots,p-1\}$, i.e. maps the set $\{1,2,\ldots,p-1\}^3$
into the set $\{0,1,2,\ldots,p-2\}$ then the left parastrophe
$(Q,\setminus)$ of a quasigroup $(Q,*)$ defined by (\ref{5}) is
defined as:
\begin{equation}
i\setminus j=\left\{ \begin{array}{cc} g(i,j,K), & \mbox{\tt If  }
g(i,j,K)\ne 0\\
p-1, & \mbox{\tt If  } g(i,j,K)=0
\end{array}
\right. \label{7}
\end{equation}
\qed
\end{corollary}

To be consistent with the notation of $f_K(j)$, we will use the
notation $g_K(i,j)$ instead the notation $g(i,j,K)$. Additional
reason for doing that will be offered in the next section, where
once the value of $K$ is chosen, it will remain fixed for
different values of $i$ and $j$.

\section{A quasigroup stream cipher in $\bbbz_p^*$}
In this section we will define a quasigroup stream cipher that
combines algorithms described in previous section. The algorithm
is as follows:

\begin{center}
{\bf A quasigroup stream cipher}
\end{center}

\begin{description}
  \item[Key generation.] Alice generates her public and private
  keys as follows:
\begin{itemize}
  \item [1.] Generate a large random prime number $p$ and a generator
  $\alpha$ of the multiplicative group $\bbbz_p^*$ of the integers
  $\{1,2,\ldots,p-1\}$.
  \item [2.] Select a random integer $a$, $1\le a \le p-2$ and compute
  $\alpha^a\ {\tt mod}\, p$
  \item [3.] Alice's public key is the triplet $(p,\alpha,\alpha^a)$;
  Alice's private key is $a$.
\end{itemize}
  \item[Session key generation.] Bob wants to establish secure
  stream channel with Alice by doing the following:
\begin{itemize}
  \item [1.] Obtain Alice's authentic public key
  $(p,\alpha,\alpha^a)$.
  \item [2.] Select a random integer $K, 1\le K \le p-1$ by which a
  quasigroup $(Q,*)$ will be defined for the elements
  $\{1,2,\ldots,p-1\}$ with equation (\ref{5}).
  \item [3.] Encrypt $K$ by ElGamal algorithm, obtaining $C=(\Gamma,\Delta)$.
  \item [4.] Select $k\ge 3$ random integers $a_i,\ i=1,2,\ldots,k,\ 1\le a_i \le p-2$ to be leaders for quasigroup stream cipher and
  encrypt them by ElGamal algorithm, obtaining $C_i=(\Gamma_i,\Delta_i),\  i=1,2,\ldots,k$.
  \item [5.] Send $C_i$.
\end{itemize}
\item[Establishment of a secure stream cipher.] Alice will
establish secure stream channel with Bob by doing the following:
\begin{itemize}
  \item [1.] Decrypt $C$ by ElGamal decryption procedure,
  obtaining $K$ by which a left parastrophe $(Q,\setminus)$ will be defined
  with equation (\ref{7}).
  \item [2.] Decrypt $C_i$ by ElGamal decryption procedure,
  obtaining the integers $a_i,\ i=1,2,\ldots,k,\ 1\le a_i \le p-1$ to be leaders for quasigroup stream
  cipher.
\end{itemize}
\item[Stream Encryption.] Bob encrypts mesages $m_{\mu}$ from the message stream $m_1,m_2,\ldots$ by doing the following:
\begin{itemize}
  \item [1.] Represent every message part $m_{\mu}$ as an integer in the range\\ $\{0,1,\ldots,p - 1\}$.
  \item [2.] Iteratively compute $m_{\mu}^{(i)}=a_i * m_{\mu}^{(i-1)}$, where $m_{\mu}^{(0)}\equiv m_{\mu}$, $i=1,\ldots,k$ and $*$ is quasigroup operation defined by (\ref{5}).
  \item [3.] Set $c_{\mu}=m_{\mu}^{(k)}$ and update the values of the leaders by $a_i=m_{\mu}^{(i)},\ i=1,\ldots,k-1$ and $a_k=1+(\sum_{i=1}^{k}m_{\mu}^{(i)})\ {\tt mod}\ (p-1)$.
  \item [4.] Send the ciphertext $c_{\mu}$.
\end{itemize}
\item[Stream Decryption.] To decrypt the part $c_{\mu}$ of the cipher text stream $c_1,c_2,\ldots$ Alice should do the
  following:
\begin{itemize}
  \item [1.] Obtain cipher text part $c_{\mu}$.
  \item [2.] Iteratively compute $c_{\mu}^{(k)}=a_k \setminus c_{\mu}$, $c_{\mu}^{(i)}=a_i \setminus c_{\mu}^{(i+1)}$, $i=k-1,\ldots,1$ and $\setminus$ is quasigroup operation defined by
  (\ref{7}).
  \item [3.] Recover $m_{\mu}=c_{\mu}^{(1)}$ and update the values of
  the leaders by $a_i=c_{\mu}^{(i+1)}$, $i=k-1,\ldots,1$ and $a_k=1+(c_{\mu}+\sum_{i=2}^{k}c_{\mu}^{(i)})\ {\tt mod}\ (p-1)$.
\end{itemize}
\end{description}

\begin{example}
In the following example, we will use relatively small value of
the prime number $p$, in order to show the work of the
algorithm.\\
\begin{description}
  \item[Key generation.] Alice generates her public and private
  keys as follows:
\begin{itemize}
  \item [1.] $p=2^{16}+1=65537$ and a generator
  $\alpha=13$ of the multiplicative group $\bbbz_p^*$ of the integers
  $\{1,2,\ldots,65536\}$.
  \item [2.] She then select a random integer $a=10307$ and compute
  $\alpha^a\ {\tt mod}\, p=13^{10307}\ {\tt mod}\, 65537=29656$
  \item [3.] Alice's public key is the triplet $(p,\alpha,\alpha^a)=(65537,13,29656)$;
  Alice's private key is $a=10307$.
\end{itemize}
  \item[Session key generation.] Bob wants to establish secure
  stream channel with Alice by doing the following:
\begin{itemize}
  \item [1.] Obtain Alice's authentic public key
  $(p,\alpha,\alpha^a)=(65537,13,29656)$.
  \item [2.] Select a random integer $K=35469$ by which a
  quasigroup $(Q,*)$ will be defined for the elements
  $\{1,2,\ldots,65536\}$ with equation\\ $i*j=\frac{i}{1+(35469+j)\,{\tt mod}\, 65536}
{\tt mod}\, 65537$.
  \item [3.] Encrypt $K$ by ElGamal algorithm, obtaining $C=(\Gamma,\Delta)=(1845,57308)$ (by using the random exponent to be $e=53882$).
  \item [4.] Select $k=3$ random integers $(a_1,a_2,a_3)=(41866,44005,27025)$ to be initial leaders for quasigroup stream cipher and
  encrypt them by ElGamal algorithm, obtaining $C_1=(\Gamma_1,\Delta_1)=(13023,32389),\ C_2=(\Gamma_2,\Delta_2)=(39691,7691)$ and
  $C_3=(\Gamma_3,\Delta_3)=(14791,21654)$ (by using random exponents to be 19495, 7737 and 4256).
  \item [5.] Send $C_1,\ C_2$ and $C_3$.
\end{itemize}
\item[Establishment of a secure stream cipher.] Alice will
establish secure stream channel with Bob by doing the following:
\begin{itemize}
  \item [1.] Decrypt $C=(\Gamma,\Delta)=(1845,57308)$ by ElGamal decryption procedure,
  obtaining $K=35469$ by which a left parastrophe $(Q,\setminus)$ will be defined
  with equation (\ref{7}) i.e. $i \setminus j =((i
\times j^{-1}\ {\tt mod}\ p) -1 -K)\ {\tt mod}\ (p-1)$.
  \item [2.] Decrypt $C_1=(\Gamma_1,\Delta_1)=(13023,32389),\ C_2=(\Gamma_2,\Delta_2)=(39691,7691)$ and
  $C_3=(\Gamma_3,\Delta_3)=(14791,21654)$ by ElGamal decryption procedure,
  obtaining the integers $(a_1,a_2,a_3)=(41866,44005,27025)$ to be initial leaders for quasigroup stream
  cipher.
\end{itemize}
\item[Stream Encryption.] Bob encrypts messages $m_{\mu}$ from the message stream $m_1,m_2,\ldots$ by doing the following:
\begin{itemize}
  \item [1.] Suppose that Bob wants to send the following three successive messages $(m_1,m_2,m_3,\ldots)=(64816,47513,52916,\ldots)$.
  \item [2.] He iteratively compute $m_1^{(1)}=a_1 * m_1=41866 * 64816 = 6851$, $m_1^{(2)}=a_2 * m_1^{(1)}=44005 * 6851 = 44908$, $m_1^{(3)}=a_3 * m_1^{(2)}=27025 * 44908=19753$.
  \item [3.] Set $c_1=m_1^{(3)}=19753$ and update the values of the leaders by $(a_1,a_2,a_3)=(m_1^{(1)},m_1^{(2)},1+(m_1^{(1)}+m_1^{(2)}+m_1^{(3)})\ {\tt mod}\ (p-1))=(6851,44908,5977)$.
  \item [4.] Send the ciphertext $c_1=19753$.
  \item [5.] He then repeats the steps 2.--4. for $m_2=47513$ and
  so on.
\end{itemize}
\item[Stream Decryption.] To decrypt the part $c_{\mu}$ of the cipher text stream $c_1,c_2,\ldots$ Alice should do the
  following:
\begin{itemize}
  \item [1.] Obtain cipher text part $c_1=19753$.
  \item [2.] Iteratively compute $c_1^{(3)}=a_3 \setminus c_1=27025  \setminus
  19753=44908$, $c_1^{(2)}=a_2 \setminus c_1^{(3)}=44005 \setminus 44908=6851$,
  $c_1^{(1)}=a_1 \setminus c_1^{(2)}= 41866 \setminus 6851=64816$.
  \item [3.] Recover $m_1=c_1^{(1)}=64816$ and update the values of
  the leaders by $a_2=c_1^{(3)}=44908, a_1=c_1^{(2)}=6851$ and $a_3=1+(c_1+c_1^{(3)}+c_1^{(2)})\ {\tt mod}\ (p-1)=1+(19753+44908+6851)\ {\tt mod}\ (p-1)=5977$.
  \item [4.] She then repeats the steps 2. and 3. for $c_2$ and so
  on.
\end{itemize}
\end{description}

\end{example}

\section{Cryptographical strength of the quasigroup stream cipher in $\bbbz_p^*$}

The proposed algorithm has two parts. The first part is the part
that is ElGamal algorithm, and the cryptographical strength of
that part is based on the strength of ElGamal algorithm i.e. on
cryptographical strength of Diffie-Helman algorithm which further
relies its security on intractability of Discrete Logarithm
Problem.

The second part is the part where fast stream cipher
transformations are performed using $k$ leaders that are unknown
for an adversary. In what follows we will examine the
cryptographical strength of the stream cipher depending on the
number of leaders $k$. We will assume that the quasigroup stream
cipher is broken if the adversary find some of the symetric parts
of the stream i.e. if he find somehow the number $K$ which defines
the permutation in $\bbbz_p^*$ or any of the initial leaders $a_1,
a_2, \ldots, a_k$.

\subsection{The case $k=1$}

Let $k=1$, and let suppose that the adversary have one pair of
known plaintext and ciphertext
$(M,C)=(m_1,m_2,m_3,\ldots,c_1,c_2,c_3,\ldots)$. By having that
information he will try to obtain some knowledge about the value
$K$ which defines the quasigroup $(Q,*)$ and about the initial
leader $a_1$. From the definition of the algorithm it follows that
$c_1=a_1 * m_1$ and $c_2=c_1*m_2$, i.e.

$$
\left\{ \begin{matrix}
c_1= & \frac{a_1}{1+(K+m_1)\,{\tt mod}\,
(p-1)}
{\tt mod}\, p\\
c_2= & \frac{c_1}{1+(K+m_2)\,{\tt mod}\, (p-1)} {\tt mod}\, p
\end{matrix}
\right.
$$
where $a_1$ and $K$ are not known. The last system can be reduced
to a quadratic polynomial equation with one unknown $K$ in the
field $\bbbz_p$. Such type of univariate quadratic polynomial
equations can be easily solved for any prime number $p$
(\cite{Cohen} p.37). So, if the number of used leaders is $k=1$
the stream cipher is easily breakable.

\subsection{The case $k=2$}
For the case when $k\ge 2$ we will make an analysis of the
strength of the algorithm by assuming that adversary can apply the
chosen plaintext attack, i.e. we will assume that the adversary
knows what is the outcome from encryption of the plaintext stream
where all messages $m_i=p-2, i=1,2,\ldots$ i.e. he knows the
following pair of plaintext and ciphertext:
$(M,C)=(p-2,p-2,p-2,p-2,\ldots,c_1,c_2,c_3,c_4\ldots)$. With that
special case, the equations for quasigroup transformations are
simplified since for any $c \in \bbbz_p^*$,
$$c*(p-2)=c \times f_K(p-2)\,{\tt mod}\, p=\frac{c}{1+(K+p-2)\,{\tt mod}\, (p-1)}\,{\tt mod}\, p=\frac{c}{K}\,{\tt mod}\, p$$

We will make an additional assumption, in order to simplify the
equations that have to be solved. Namely, instead of complicated
usage of modulo $p$ and modulo $p-1$ in the obtained equations, we
will only use operations modulo $p$. Although the solutions for
those equations are not necessary solutions for the real equations
involving modulo $p$ and modulo $p-1$, we will show that even
those simplified equations are hard to solve if the number of used
leaders $k$ is sufficiently large.

So, by mentioned simplifications and assumptions, for $k=2$ the
adversary will obtain the following system of equations in
$\bbbz_p$:

$$
\left\{ \begin{matrix}
c_1 = & \frac{a_2}{1 + \frac{a_1}{K} + K}\\
c_2=  & \frac{c_1 + \frac{a_1}{K}}{1 + \frac{a_1}{K^2} + K}\\
c_3=  & \frac{c_2 + \frac{a_1}{K^2}}{1 + \frac{a_1}{K^3} + K}
\end{matrix}
\right.
$$

The last system can be reduced to the following univariate
polynomial equation of degree 3 in $\bbbz_p$ with unknown variable
$K$:
$$c_3\,K^3 + \left( -2\,c_2 + c_3 - c_2\,c_3 \right) \,K^2 + \left( c_1 - c_2 + {c_2}^2 \right) \,K +
c_2\,c_3 - c_1\,c_3 = 0$$

For those type of polynomials there are efficient (running in
polynomial time) randomized algorithms for solving them in
$\bbbz_p^*$ (see for example \cite{Cohen} p.37, p.123-p.132).

So, we could say that the case with two leaders i.e. when $k=2$
when the equations are simplified and we only work modulo $p$, can
be successfully attacked by the chosen plaintext attack.

We are not aware if there are some known randomized or
deterministic algorithms for solving equations that involve both
modulo $p$ and modulo $p-1$ which is much complicated and harder
to solve case, but taking conservative approach, we will consider
that the case $k=2$ is not safe.

\subsection{The case $k=3$}
For the case when $k=3$, and by supposing that a possible
adversary have one pair of known chosen plaintext and ciphertext
$(M,C)=(p-2,p-2,p-2,p-2,\ldots,c_1,c_2,c_3,c_4\ldots)$, he can
obtain the following system of simplified equations:

\begin{equation*}
\left\{ \begin{array}{l} \vspace{0.4cm} c_1=\frac{a_3}{1 + K +
\frac{a_2}{1 +
\frac{a_1}{K} + K}}\\
\vspace{0.4cm} c_2=\frac{c_1 + \frac{a_1}{K} + \frac{a_2}{1 +
\frac{a_1}{K} + K}}
   {1 + K + \frac{a_2}{\left( 1 + \frac{a_1}{K^2} + K \right) \,\left( 1 + \frac{a_1}{K} + K \right)
   }}\\
   \vspace{0.4cm}
c_3=\frac{c_2 + \frac{a_1}{K^2} + \frac{a_2}
      {\left( 1 + \frac{a_1}{K^2} + K \right) \,\left( 1 + \frac{a_1}{K} + K \right) }}{1 + K +
     \frac{a_2}{\left( 1 + \frac{a_1}{K^3} + K \right) \,\left( 1 + \frac{a_1}{K^2} + K \right) \,\left( 1 + \frac{a_1}{K} + K \right)
     }}\\
     \vspace{0.4cm}
c_4 =\frac{c_3 + \frac{a_1}{K^3} + \frac{a_2}
      {\left( 1 + \frac{a_1}{K^3} + K \right) \,\left( 1 + \frac{a_1}{K^2} + K \right) \,\left( 1 + \frac{a_1}{K} + K \right) }}{1 + K +
     \frac{a_2}{\left( 1 + \frac{a_1}{K^4} + K \right) \,\left( 1 + \frac{a_1}{K^3} + K \right) \,
        \left( 1 + \frac{a_1}{K^2} + K \right) \,\left( 1 + \frac{a_1}{K} + K \right) }}
\end{array}
\right.
\end{equation*}

If we introduce two new variables $A_1=\frac{a_1}{K}$ and
$A_2=\frac{a_2}{1 + \frac{a_1}{K} + K}$ we can reduce the above
system to the system of two bivariate polynomial equations:
\begin{equation*}
\left\{ \begin{array}{l} P_1(K,A_1) = 0 \\ P_2(K,A_1) = 0
\end{array}
\right.
\end{equation*}
where in the first polynomial $P_1$, the degree of $K$ is 7, and
the degree of $A_1$ is 3, and in the second polynomial $P_2$, the
degree of $K$ is 12 and the degree of $A_1$ is 5.

It is clear that if we continue several more steps, with a usage
of several more leaders, the complexity of the system to be solved
would increase even more. Although the obtained equations have
specific structure we can ask the following question: Are there
any fast (in polynomial time, deterministic or randomized)
algorithms for solving systems of multivariate polynomials modulo
big prime number $p$. We can try to find the answer in the results
of modern Number Theory. Namely, two areas of research are
connected with posted question: 1. Factorization of multivariate
polynomials modulo prime number and 2. Solving systems of
multivariate polynomials modulo prime number. Although in the last
decades we see dramatic breakthrough in factorization of
multivariate polynomials modulo prime numbers (see for example
\cite{Gathen}, \cite{Lenstra85}, \cite{Gao}, \cite{Yo02} and
\cite{Bostan}), and in some cases the results of that breakthrough
increase our knowledge how to solve systems of multivariate
polynomials modulo prime number, it was shown that finding the
roots of systems of multivariate polynomials modulo big prime
number $p$ is equivalent to solving an NP-hard problem (see
\cite{Kaltofen} and \cite{Plaisted}).

From above discussion, we can say that we have strong evidence
that breaking the proposed stream cipher would be as hard as
solving in polynomial time some NP-hard problem.

\section{Conclusions and further directions}
In these conclusions, we would like to say something about the
speed of the proposed stream cipher. For every block $m_\mu$ and
$c_\mu$ both in encryption and decryption phase $k$ modular
multiplications and $k$ modular calculations of inverse element
are needed, but doesn't need operations of modular exponentiation.
If we have in mind that modular multiplication and modular
division operations modulo $p$ can be implemented in $O(\log_2^2
p)$, that means that total number of operations have complexity of
$O(k \log_2^2 p)$. In other words, calculated as operations per
byte, the proposed stream cipher is much faster then cryptographic
algorithms that work over $\bbbz_p^*$, and can approach the speed
of fast symmetric-key stream ciphers. However, the benefits for
using the proposed stream cipher are that its cryptographic
strength is equivalent as solving in polynomial time (with
deterministic or with randomized algorithm) NP-hard problems.

From other point of view, the proposed stream cipher tries to
bridge the gap between fast symmetric-key algorithms and slow
public-key algorithms, using the flexibility of key-exchange
possibilities of the public-key algorithms, and the speed of
symmetric-key algorithms. The mathematical structure of the domain
of encoded messages is the same in both parts, i.e. the
transformations are done in the set of $\bbbz_p^*=\bbbz_p
\setminus \{0\}=\{1,2,...,p-1\}$.

In practical implementation, in order to avoid the common
disadvantage of all public-key algorithms that is the expansion of
the original message first in the process of transformation of a
message $m$ into an integer from the set $\{1,2,\ldots,p-1\}$ and
then in the process of encryption, we should implement the
proposed algorithm with a prime number $p$ which has the form of a
Fermat prime number $F_n=2^{2^n}+1$. However, for $n=3$, and $n=4$
the prime numbers $F_3=257$ and $F_4=65537$ are too small for
cryptographic purposes, and there are no prime Fermat numbers for
$n > 4$. To overcome that disadvantage we propose the use of a
prime numbers of the form $p_l=2^{8 l}+3$. For example $p_{98}$,
$p_{213}$, and $p_{251}$ are prime numbers with 784, 1704 and 2008
bits respectfully. For example let suppose that we use a prime
number $p_{251}$ with 2008 bits. Then, in the process of message
transformation, we could simply treat every consecutive 2008 bits
i.e. 251 bytes as an input message and to add one extra byte that
will be in fact the total message expansion.

\vspace{0.9cm} {\bf ACKNOWLEDGMENT} \vspace{0.5cm}

I would like to thank prof. Kaltofen for his suggestions how to
improve some parts of this paper.

%
%

\end{document}